\documentstyle[12pt]{article}
  
\begin{document}
\newcommand{\mysection}[1]{\setcounter{equation}{0}\section{#1}}

% Journals
\newcommand{\jref}[4]{{\it #1} {\bf #2}, #3 (#4)}
\newcommand{\NPB}[3]{\jref{Nucl.\ Phys.}{B#1}{#2}{#3}}
\newcommand{\PLB}[3]{\jref{Phys.\ Lett.}{#1B}{#2}{#3}}
\newcommand{\PR}[3]{\jref{Phys.\ Rep.}{#1}{#2}{#3}}
\newcommand{\PRD}[3]{\jref{Phys.\ Rev.}{D#1}{#2}{#3}}
\newcommand{\PRL}[3]{\jref{Phys.\ Rev.\ Lett.}{#1}{#2}{#3}}
\newcommand{\PRV}[3]{\jref{Phys.\ Rev.}{#1}{#2}{#3}}

% draw box with width #1pt and line thickness #2pt
\newcommand{\drawsquare}[2]{\hbox{%
\rule{#2pt}{#1pt}\hskip-#2pt%  left vertical
\rule{#1pt}{#2pt}\hskip-#1pt%  lower horizontal
\rule[#1pt]{#1pt}{#2pt}}\rule[#1pt]{#2pt}{#2pt}\hskip-#2pt%  upper horizontal
\rule{#2pt}{#1pt}}% right vertical

% Young tableaux
\newcommand{\Yfund}{\raisebox{-.5pt}{\drawsquare{6.5}{0.4}}}%  fund
\newcommand{\Ysymm}{\raisebox{-.5pt}{\drawsquare{6.5}{0.4}}\hskip-0.4pt%
        \raisebox{-.5pt}{\drawsquare{6.5}{0.4}}}%  symmetric second rank
\newcommand{\Yasymm}{\raisebox{-3.5pt}{\drawsquare{6.5}{0.4}}\hskip-6.9pt%
        \raisebox{3pt}{\drawsquare{6.5}{0.4}}}%  antisymmetric second rank
\newcommand{\Ythree}{\raisebox{-3.5pt}{\drawsquare{6.5}{0.4}}\hskip-6.9pt%
        \raisebox{3pt}{\drawsquare{6.5}{0.4}}\hskip-6.9pt
        \raisebox{9.5pt}{\drawsquare{6.5}{0.4}}}

\begin{titlepage}
\begin{center}
{\hbox to\hsize{hep-th/9704067 \hfill  MIT-CTP-2622}}
\bigskip

\bigskip

{\Large \bf  $N=1$ Supersymmetric Product Group Theories
in the Coulomb Phase}

\bigskip

\bigskip

{\bf Csaba Cs\'aki$^{a,}$\footnote{Address after August 1: 
Department of Physics, 
University of California, Berkeley, CA 94720.}, Joshua Erlich$^a$, Daniel Freedman$^{a,b}$ \\ and
Witold Skiba$^{a,}$\footnote{Address after September 1: Department of Physics, 
University of California at San Diego, La Jolla  CA 92093.} }\\

\bigskip

{ \small \it $^a$Center for Theoretical Physics,

Massachusetts Institute of Technology, Cambridge, MA 02139, USA }

\bigskip

{ \small \it $^b$Department of Mathematics,
Massachusetts Institute of Technology, Cambridge, MA 02139, USA }

{\tt csaki@mit.edu, jerlich@ctp.mit.edu, dzf@math.mit.edu, 
skiba@mit.edu}

\bigskip

\vspace*{1cm}
{\bf Abstract}\\
\end{center}

\noindent
We study the low-energy behavior of $N=1$ supersymmetric gauge theories with 
product gauge groups $SU(N)^M$ and $M$ chiral superfields 
transforming in the fundamental
representation of two of the $SU(N)$ factors.  These theories are in
the Coulomb phase with an unbroken $U(1)^{N-1}$ gauge group.  For
$N \geq 3$, $M \geq 3$ the theories are chiral. The low-energy gauge
kinetic functions can
be obtained from hyperelliptic curves which we derive by considering
various limits of the theories. We present several consistency checks of
the curves including confinement through the addition of mass perturbations 
and other limits.

\bigskip

\bigskip

\end{titlepage}

\section{Introduction}

There has been a dramatic progress in our understanding
of the dynamics of supersymmetric gauge theories 
during the past three years. Seiberg and Witten
gave a complete solution to the low-energy
dynamics of $N=2$ supersymmetric $SU(2)$ theory
with or without fundamental matter fields~\cite{SW}.
This work has been generalized to 
pure $N=2$ $SU(N)$ theories with and without fundamental matter fields as well
as to other gauge groups~\cite{ArgyresKlemm,Argyres,pureSOSp,fundamental}.

Following Seiberg's work on $N=1$ supersymmetric QCD~\cite{Seiberg},
there is a growing number of exact results in $N=1$ theories 
as well~\cite{phases,SO,ILS,intin,N=1}. However in these theories 
one does not have a 
complete solution of the low-energy dynamics, but only the
exact form of the superpotential. The major difference between
$N=2$ and $N=1$ theories is that in $N=2$ the full Lagrangian is determined
in terms of a holomorphic prepotential, while in $N=1$ the
superpotential and the gauge-kinetic term are holomorphic,
but the K\"ahler potential is not. 

Intriligator and Seiberg noted that the methods which are used to solve
certain $N=2$ theories can also be applied to Coulomb branches of $N=1$
theories~\cite{phases}. In the Coulomb phase there are massless photons
in the low-energy theory, whose couplings to the matter fields are described
by the following Lagrangian:
\begin{displaymath}
 L=\frac{1}{4 \pi} {\rm Im}
 \int d^2 \theta \tau_{ij} W_{\alpha}^i
    W^{j \alpha},
\end{displaymath}
where $W_{\alpha}^i$ is the field strength chiral superfield, corresponding
to the $i$th $U(1)$ factor and $\tau_{ij}$ is the effective
gauge coupling, which is a holomorphic function of the matter fields.
Often this $\tau_{ij}$ can be identified with the period matrix
of a hyperelliptic curve. Thus for theories in the Coulomb phase,
an important part of 
the solution of the low-energy dynamics can be found by determining
the hyperelliptic curve as a function of the moduli and the
scales of the theory. The singular points of the curve usually
signal the existence of massless monopole or dyon superfields, whose
properties can be read off from the curve.

Except for the $N=2$ theories based on $SU$, $Sp$ and $SO$ groups with matter
fields in the fundamental 
representation~\cite{SW,ArgyresKlemm,pureSOSp,fundamental}, 
there are very few theories
for which the description of the Coulomb branch is known. The other
examples include $N=2$ $G_2$ theory with no matter fields~\cite{G2};
$N=1$ $SU$, $Sp$ and $SO$ theories with adjoint and fundamental matter and
a Landau-Ginsburg type superpotential~\cite{LG};
and also $N=1$ $SO(M)$ theories with
$M-2$ vectors~\cite{SO}.

In this paper we examine $N=1$ theories with product gauge groups $SU(N)^M$ and
$M$ matter fields, each transforming as a fundamental under exactly
two $SU(N)$ factors. All of these theories are in the Coulomb phase.
The $SU(N)^M$ theory contains an unbroken $U(1)^{N-1}$ gauge group.
For each of these theories we identify the independent gauge invariant
operators, which parameterize the moduli space. We determine the
hyperelliptic curves describing the gauge coupling function by considering
different limits in which the theory has to reproduce known results
for other theories. We give several consistency checks for these curves.
The theories where $N\geq 3$, $M\geq 3$ are the first examples
of chiral theories in the Coulomb phase; thus, one might hope that they will 
be useful for building models of dynamical supersymmetry breaking.

The paper is organized as follows. In the next section we first review
the $SU(2) \times SU(2)$ theory of Intriligator and Seiberg~\cite{phases} and 
then generalize this theory to $SU(2)^N$. We explain the
$SU(2)^3$ case in detail and show that the singularities produce the
expected behavior when the theory is perturbed by adding mass terms.
Section~\ref{sec:SUN2} describes the $SU(N)\times SU(N)$ 
theories, while curves
for the general $SU(N)^M$ theories are given in 
Section~\ref{sec:SUNM}. We conclude
in Section~\ref{sec:conclusions}. 
An appendix contains an analysis of 
the D-flat conditions in the general $SU(N)^M$ theories.

%%%%%%%%%%%%%%%%%%%%%%%%%%%%%%%%%%%%%%%%%%%%%%%%%%%%%%%%%%%%%%%%%%%%%%
\section{$SU(2)^N$\label{sec:SU2N}}
%%%%%%%%%%%%%%%%%%%%%%%%%%%%%%%%%%%%%%%%%%%%%%%%%%%%%%%%%%%%%%%%%%%%%%

In this section we first review the pure $N=2$ $SU(N)$ theories and the 
$SU(2)\times SU(2)$ theory of Intriligator and Seiberg~\cite{phases}. Then
we generalize this $SU(2)\times SU(2)$ theory to $SU(2)^N$ . 

The hyperelliptic curves for pure $N=2$ $SU(N)$ theories were given in
Ref.~\cite{ArgyresKlemm}. This solution can be summarized as
follows. The moduli space of the Coulomb branch can be
parameterized by the expectation values of the
independent gauge invariant operators formed from the
adjoint field $\Phi$:
\[ u_k=\frac{1}{k}{\rm Tr}\, \Phi^k,\, k>1.\]
The expectation value of the adjoint can always be rotated 
to a diagonal form
\[ \Phi =\left( \begin{array}{cccc} a_1& \\
&a_2& \\
& & \ddots \\
& & & a_N\end{array} \right), \; \; \sum a_i=0,\]
where classically $u_k=\frac{1}{k}\sum_{i=1}^N a_i^k$. 
It was argued in Ref.~\cite{ArgyresKlemm} that the $N=2$ pure
$SU(N)$ Yang-Mills theory can be described in terms of a genus
$N-1$ Riemann surface. The hyperelliptic 
curve describing this surface
is given by
\begin{equation}
\label{suncurve}
y^2=\Pi_{i=1}^{N}(x-a_i)^2-4\Lambda^{2N},
\end{equation}
where $\Lambda$ is the dynamical scale of the $SU(N)$ theory and products of
$a_i$'s are to be written in terms of the $u_k$.
In terms of the
variables $s_k$, which are defined in the classical regime by \begin{equation}
s_k=(-1)^k\, \sum_{i_1<\ldots  <i_k} a_{i_1}\ldots a_{i_k},\ k=2,\ldots ,N,
\label{eq:s=a}
\end{equation}
this curve can also be conveniently expressed as
\begin{equation}
\label{suncurve2}
y^2=(x^N+\sum_{i=2}^{N}s_ix^{N-i})^2-4\Lambda^{2N}.
\end{equation}
The variables $s_k$ are related to the $u_k$'s by Newton's 
formula, with $s_0=1$ and $s_1=u_1=0$,
\begin{equation} ks_k+\sum_{j=1}^k js_{k-j}u_j=0, \label{eq:newton}
\end{equation}
thus defining them quantum mechanically.

Intriligator and Seiberg pointed out~\cite{phases} that the
techniques used for solving $N=2$ theories can be applied to the 
Coulomb branches of $N=1$ theories as well. However, in this 
case the determination of $\tau$ does not imply a complete solution
of the theory. Intriligator and Seiberg showed several examples
where the gauge coupling $\tau$ can be exactly 
determined. Their result for the $SU(2)\times SU(2)$ theory
with $2(\Yfund ,\Yfund )$ can be summarized as follows. 

The field content of the $SU(2)\times SU(2)$ theory is 
$(Q_i)_{\alpha \beta}$, where $i$ is the flavor index and
$\alpha ,\beta$ are the $SU(2)$ indices. The three 
independent gauge 
invariant operators are 
\begin{equation}
\label{m's}
M_{ij}=\frac{1}{2} (Q_i)_{\alpha \beta}
(Q_j)_{\alpha ' \beta '}\epsilon^{\alpha \alpha '}
\epsilon^{\beta  \beta '}.\nonumber
\end{equation}
On a generic point of the moduli space the $SU(2)\times SU(2)$
gauge symmetry is broken to $U(1)$; thus the theory is in an
Abelian Coulomb phase. It is natural to assume that the
Coulomb phase can be described by a genus one Riemann surface
determined by an elliptic curve, where the coefficients of 
$x$ are functions of the scales $\Lambda_{1,2}$ and the 
moduli $M_{ij}$. This curve can be  determined by considering 
two different limits of the theory. One limit involves breaking
the $SU(2)\times SU(2)$ to the diagonal $SU(2)$ group by
giving a diagonal VEV to $Q_1$, while the other limit is
$\Lambda_2\gg \Lambda_1$, where $SU(2)_2$ is confining with a quantum modified
constraint~\cite{Seiberg}. In both limits the theory reduces to an 
$SU(2)$ theory with an adjoint chiral superfield, whose 
elliptic curve is given in Eq.~\ref{suncurve2}. These two limits
completely fix the genus one elliptic curve, whose fourth order
form is given by
\begin{equation}
\label{su2su2}
y^2=(x^2-(U-\Lambda_1^4-\Lambda_2^4))^2-4\Lambda_1^4\Lambda_2^4,
\end{equation}
where $U={\rm det}\,M$.
Note that the form of the curve is just what we would get
for an $N=2$ $SU(2)$ theory, except that the modulus
$U$ (which is to be thought of as a function of the $M$'s) 
is shifted by a constant, and that the scale is the product of the scales
of each $SU(2)$ factor.  This scale is
determined by matching to the diagonal theory. A similar situation will hold 
for the more general $SU(N)\times SU(N)$ theories, and with the help
of these curves we will be able to describe a
general class of $SU(N)^M$ theories as well.

Now we generalize the $SU(2)\times SU(2)$ theory
of Intriligator and Seiberg~\cite{phases} presented above 
to theories based on the
$SU(2)_1 \times SU(2)_2 \times \ldots \times SU(2)_N$ product group.
The field content of the theory is described in the table below:
\begin{equation}
 \begin{array}{c|ccccc}
      & SU(2)_1 & SU(2)_2 & SU(2)_3 & \ldots & SU(2)_N \\ \hline
  Q_1 & \Yfund  & \Yfund  & 1       & \ldots & 1 \\
  Q_2 & 1       & \Yfund  & \Yfund  & \ldots & 1 \\
  \vdots & \vdots & \vdots & \vdots & \vdots & \vdots \\
  Q_N & \Yfund & 1 & 1 & \ldots & \Yfund 
 \end{array}\;. \label{eq:fields}
\end{equation}
The classical moduli space of this theory can be parameterized by the
following gauge invariants
\begin{eqnarray}
M_i={\rm det}\,Q_i=\frac{1}{2} (Q_i)_{\alpha_i \alpha_{i+1}} 
(Q_i)_{\beta_i \beta_{i+1}}
    \epsilon^{\alpha_i \beta_i} \epsilon^{\alpha_{i+1} \beta_{i+1}},
    \; \; i=1,\ldots,N \nonumber \\
T=\frac{1}{2} (Q_1)_{\beta_1 \alpha_2} (Q_2)_{\beta_2 \alpha_3}
  (Q_3)_{\beta_3 \alpha_4} \ldots (Q_1)_{\beta_N \alpha_1}
  \epsilon^{\alpha_1 \beta_1} \epsilon^{\alpha_2 \beta_2} \ldots
  \epsilon^{\alpha_N \beta_N}.  
\end{eqnarray}
As shown in the appendix, 
generic vacuum expectation values of these operators preserve a $U(1)$
gauge symmetry. We will describe the behavior of the holomorphic gauge
coupling for the $U(1)$ gauge group by constructing an elliptic curve.

We first analyze the   $SU(2)_1 \times SU(2)_2 \times SU(2)_3$ theory,
which can be reduced in various limits to the $SU(2)_1 \times SU(2)_2$ theory
of Ref.~\cite{phases}. By exploring the limit of large VEV for the
field $Q_3$ and the limit $\Lambda_3 \gg \Lambda_1,\ \Lambda_2$
we will be able to determine the coefficients of the curve. We will
work with the third order form of the elliptic curve, since that form
is more convenient in this case. First consider the limit of large diagonal
VEV, $v$, for $Q_3$. In this limit $SU(2)_1 \times SU(2)_3$ is broken
to its diagonal subgroup $SU(2)_D$. Three components of $Q_3$ are eaten
by the Higgs mechanism, while the remaining component is a singlet
of $SU(2)_D$. Both $Q_1$ and $Q_2$ 
transform as $(\Yfund,\Yfund)$
under the unbroken $SU(2)_D \times SU(2)_2$.

This $SU(2)\times SU(2)$ theory is precisely the theory of
Ref.~\cite{phases} described above. The invariants of this theory are
$\tilde{M}_{11}=Q_1 Q_1$, $\tilde{M}_{22}=Q_2 Q_2$ and 
$\tilde{M}_{12}=Q_1 Q_2$, which can be expressed in terms of the
invariants of the original $SU(2)_1\times SU(2)_2\times SU(2)_3$ theory:
\begin{displaymath}
M_1= \tilde{M}_{11},\; \;  M_2= \tilde{M}_{22},\; \;  T=\tilde{M}_{12} v,\; 
\; {\rm and} \; \; M_3=v^2. 
\end{displaymath}
The third order curve for the $SU(2)_D \times SU(2)_2$ theory is
\begin{equation}
\label{eq:basiccurve}
 y^2 = x^3 + x^2 (\Lambda_D^4 + \Lambda^4_2 - \tilde{M}_{11} \tilde{M}_{22}
                 + \tilde{M}_{12}^2) + x  \Lambda_D^4 \Lambda^4_2,
\end{equation}
where $\Lambda_D^4 = \Lambda_1^4 \Lambda_3^4 / M_3^2$. 

Let us express this limit of the elliptic curve in terms of the original
gauge invariants:
\begin{equation}
 y^2 = x^3 + x^2 ( \frac{\Lambda_1^4 \Lambda^4_3}{M_3^2} + \Lambda_2^4 -
 M_1 M_2 + \frac{T^2}{M_3} ) + x  \frac{\Lambda_1^4 \Lambda_2^4 \Lambda_3^4}{
 M_3^2}. \nonumber 
\end{equation}
After rescaling the above curve by $x\rightarrow x / M_3$, $y \rightarrow
y / M_3^{3/2}$ we obtain
\begin{equation}
 y^2 = x^3 + x^2 ( \frac{\Lambda_1^4 \Lambda^4_3}{M_3^2} + \Lambda_2^4 M_3 -
 M_1 M_2 M_3 + T^2 ) + x \Lambda_1^4 \Lambda_2^4 \Lambda_3^4.\label{eq:10}
\end{equation}
Since this curve is only valid in the limit of large $v$, the term
$x^2 \Lambda_1^4 \Lambda_3^4 /M_3^2$ is of lower order than other terms
proportional to $x^2$, and should be neglected. 

The final form of the
curve has to be invariant under all symmetries of the theory. For
instance, simultaneous interchange of $\Lambda_1$ with $\Lambda_2$
and interchange of $Q_2$ with $Q_3$ does not change the theory, and there are
other similar permutations.  The only term that is not invariant
under such permutations is $\Lambda_2^4 M_3$. The properly symmetrized 
combination
is of the form $\Lambda_1^4 M_2+ \Lambda_2^4 M_3+\Lambda_3^4 M_1$.
The final expression for the curve is therefore
\begin{equation}
\label{su2cube}
  y^2=x^3 + x^2 (\Lambda_1^4 M_2 + \Lambda_2^4 M_3 + \Lambda_3^4 M_1 -
      M_1 M_2 M_3 + T^2) + x  \Lambda_1^4 \Lambda_2^4 \Lambda_3^4\, ,
\end{equation}
while the equivalent quartic form is \begin{equation}
y^2=(x^2-(\Lambda_1^4 M_2 + \Lambda_2^4 M_3 + \Lambda_3^4 M_1 -
      M_1 M_2 M_3 + T^2))^2-4\Lambda_1^4\Lambda_2^4\Lambda_3^4\,.
\label{eq:4th}\end{equation}
It turns out that this is the complete form of the elliptic curve
for the $SU(2)^3$ theory. All other terms consistent with the symmetries,
such as $xT^6$, are excluded by the requirement of agreement with
Eq.~\ref{eq:10} in the limit of large VEV for $Q_3$.

We will present consistency checks which support our claim that the curve
derived in the large VEV limit is indeed correct. First, let us consider
the theory in the 
limit $\Lambda_3 \gg \Lambda_1,\ \Lambda_2$. The $SU(2)_3$ theory
has the same number of flavors as the number of colors. Below $\Lambda_3$,
the $SU(2)_3$ group is confining and we need to express the degrees of freedom
in terms of confined fields subject to the quantum modified
constraint~\cite{Seiberg}. The confined fields are $(Q_2^2)$, $(Q_3^2)$
and the $2\times 2$ matrix $(Q_2 Q_3)$. The quantum modified constraint in an 
$SU(2)$ theory with four doublets, ${\rm Pf}\, (q_i q_j) = \Lambda^4$,
when written in terms of the
fields confined by the $SU(2)_3$ dynamics is
\begin{displaymath}
  \Lambda_3^4= (Q_2^2) (Q_3^2) - (Q_2 Q_3)^2.
\end{displaymath}
Again, we can express invariants of the effective $SU(2)_1\times SU(2)_2$
theory in terms of $SU(2)^3$ invariants: $\tilde{M}_{11}=Q_1^2=M_1$,
$\mu \tilde{M}_{12}=Q_1 (Q_2 Q_3) =T$ and $\mu^2 \tilde{M}_{22}=(Q_2 Q_3)^2 =
M_2 M_3 - \Lambda_3^4$, where the last equality makes use of the quantum
modified constraint. The factors of dimensional constant $\mu$ are included
in order to make the $\tilde{M}$'s dimension two.

The elliptic curve for the $SU(2)_1 \times SU(2)_2$ theory is the same as in
Eq.~\ref{eq:basiccurve}, except for the obvious substitution $\Lambda_D
\rightarrow \Lambda_1$. In terms of $SU(2)^3$ invariants the curve is
\begin{displaymath}
  y^2=x^3 + x^2 \left(\Lambda_1^4 + \Lambda_2^4 - M_1
                 \frac{M_2 M_3 -\Lambda_3^4}{\mu^2} + \frac{T^2}{\mu^2}\right)
      + x \Lambda_1^4 \Lambda_2^4.
\end{displaymath}
After rescaling $x \rightarrow x/\mu^2$ and $y \rightarrow y/\mu^3$ we obtain
\begin{equation}
  y^2=x^3 + x^2 \left(\Lambda_1^4 \mu^2 + \Lambda_2^4 \mu ^2 + M_1 \Lambda_3^4
                 - M_1 M_2 M_3 + T^2\right) + x \mu^4 \Lambda_1^4 \Lambda_2^4.
\label{eq:mu}\end{equation}      
The symmetrized form of Eq.~\ref{eq:mu} with $\mu=\Lambda_3$ is identical to
Eq.~\ref{su2cube} up to the irrelevant subdominant term $x^2(\Lambda_1^4+
\Lambda_2^4)\Lambda_3^2$.

As another  consistency check  we consider integrating out all
matter fields from the $SU(2)^3$ theory. This way we obtain
three decoupled pure $SU(2)$ Yang-Mills theories whose low-energy
behavior is known and should be reproduced by the above 
description of the theory.

In order to integrate out the matter fields we add a
tree-level superpotential 
\begin{equation}
  W_{tree}=m_1M_1+m_2M_2+m_3M_3\nonumber
\end{equation}
to the theory, which corresponds to adding 
mass terms for all $Q_i$
fields.  On the singular manifold of the curve there are
massless monopoles or dyons which have to be included 
into the low-energy effective superpotential.
The curve described by Eq.~\ref{su2cube} is singular when
\begin{equation}
\label{sing}
 -T^2+M_1M_2M_3-\Lambda_1^4M_2-\Lambda_2^4M_3-\Lambda_3^4M_1=
 \pm 2\Lambda_1^2\Lambda_2^2\Lambda_3^2.
\end{equation}
Thus the low-energy effective superpotential is given by
\begin{eqnarray}
 W=&&(-T^2+M_1M_2M_3-\Lambda_1^4M_2-\Lambda_2^4M_3-\Lambda_3^4M_1
 + 2\Lambda_1^2\Lambda_2^2\Lambda_3^2)\tilde{E}_{+}E_{+}
 \nonumber \\
 &&+
 (-T^2+M_1M_2M_3-\Lambda_1^4M_2-\Lambda_2^4M_3-\Lambda_3^4M_1
 -2\Lambda_1^2\Lambda_2^2\Lambda_3^2)\tilde{E}_{-}E_{-}
 \nonumber \\
 &&+m_1M_1+m_2M_2+m_3M_3,\nonumber
\end{eqnarray}
where $\tilde{E}_{+}$ and $E_{+}$ are the superfields
corresponding
to the massless mo\-no\-poles at the first singular manifold,
while $\tilde{E}_{-}$ and $E_{-}$ are the dyons which are
massless at the second singular manifold.
The equations of
motion with respect to the fields $T,M_i,\tilde{E}_{\pm},E_{\pm}$
will determine the possible vacua of the theory.

The $M_i$ equations require that either $\tilde{E}_{+}E_{+}$
or $\tilde{E}_{-}E_{-}$ is non-vanishing, which together 
with the $\tilde{E}_{\pm}$ equations will fix the solutions
to be on one of the singular submanifolds. The $T$ equation sets
$T$ to zero and thus we are left with the following set of
equations:
\begin{eqnarray}
\label{vacua}
 &&M_1M_2M_3 -\Lambda_1^4M_2-\Lambda_2^4M_3-\Lambda_3^4M_1\pm
 2\Lambda_1^2\Lambda_2^2\Lambda_3^2 =0 \nonumber \\
 &&(M_1M_2 -\Lambda_2^4) e +m_3=0 \nonumber \\
 &&(M_1M_3 -\Lambda_1^4) e +m_2=0 \nonumber \\
 &&(M_2M_3 -\Lambda_3^4) e +m_1=0, 
\end{eqnarray}
where $e$ is the value of the monopole condensate $\tilde{E}E$.
One can show that there are eight solutions to these equations which
reproduce the  vacua
obtained from gaugino condensation which we now derive.

For large $m_i$ the $Q_i$ fields can be integrated out, and the resulting
theory consists of three decoupled pure $SU(2)$ Yang-Mills theories with
scales determined by matching:
\begin{displaymath}
 \tilde{\Lambda}_1^6=m_1m_3\Lambda_1^4,\; \;
 \tilde{\Lambda}_2^6=m_1m_2\Lambda_2^4,\; \;
 {\rm and}\; \; \tilde{\Lambda}_3^6=m_2m_3\Lambda_3^4.
\end{displaymath}
Gaugino condensation is then expected to produce a low-energy
superpotential
\begin{eqnarray}
\label{gaugino}
 W&=&2\epsilon_1 \tilde{\Lambda}_1^3+2\epsilon_2 
\tilde{\Lambda}_2^3+2\epsilon_3 
\tilde{\Lambda}_3^3 \nonumber \\
 &=& 2\epsilon_1 \sqrt{m_1m_3}\Lambda_1^2+
 2\epsilon_2 \sqrt{m_1m_2}\Lambda_2^2+
 2\epsilon_3 \sqrt{m_2m_3}\Lambda_3^2,
\end{eqnarray}
where $\epsilon_i=\pm 1$. Since the masses $m_i$ can be viewed
as source terms for the gauge invariant operators $M_i$,
the VEV's of the gauge invariants are determined by~\cite{intin}
$\frac{\partial W}{\partial m_i}=\langle M_i \rangle$, 
$\langle T \rangle =0$. The resulting vacua 
\begin{eqnarray}
\label{vac}
&& \langle M_1 \rangle = \epsilon_2 \sqrt{\frac{m_3}{m_1}} 
\Lambda_1^2 +\epsilon_3 \sqrt{\frac{m_2}{m_1}} 
\Lambda_2^2 \nonumber \\
&&\langle M_2 \rangle = \epsilon_1 \sqrt{\frac{m_1}{m_2}} 
\Lambda_2^2 +\epsilon_3 \sqrt{\frac{m_3}{m_2}} 
\Lambda_3^2 \nonumber \\
&&\langle M_3 \rangle = \epsilon_1 \sqrt{\frac{m_1}{m_3}} 
\Lambda_1^2 +\epsilon_2 \sqrt{\frac{m_2}{m_3}} 
\Lambda_3^2 \nonumber \\
\end{eqnarray}
can be shown to exactly coincide with the solutions of 
Eqs.~\ref{vacua}, providing us with a non-trivial 
check on the consistency of the curve for the $SU(2)^3$ theory.

It is quite straightforward to generalize the $SU(2)^3$ curve to $SU(2)^N$
theories with the matter content given in Table~\ref{eq:fields}.
We proceed as before and determine the curves from the limit of
large diagonal VEV for one of the $Q_i$'s and the limit in which one
of the $SU(2)$'s becomes strong. The resulting curve is:

\begin{equation}
  y^2=x^3 + x^2 \left( T^2-\prod_i M_i+(M_iM_{i+1}\rightarrow -\Lambda_{i+1}^4) \right) + x \prod_{i} \Lambda_i^4, \label{su2^N}
\end{equation}
where the last term in parentheses proportional to $x^2$ follows by substituting any 
set of nearest
neighbor bilinears $M_iM_{i+1}$ in $\prod_i M_i$ by the dynamical scale $-\Lambda_{i+1}^4$
of the 
common gauge group that both
$Q_i$ and $Q_{i+1}$ transform under.  This is a consequence of the 
quantum modified constraint in the strong coupling limit in the i+1 gauge 
group. For example, for the case with four $SU(2)$ factors
the term proportional to $x^2$ is $(T^2-M_1M_2M_3M_4+\Lambda_1^4 M_2M_3
+\Lambda_2^4 M_3M_4+\Lambda_3^4 M_4M_1+\Lambda_4^4 M_1M_2 -\Lambda_1^4\Lambda_3^4
-\Lambda_2^4\Lambda_3^4)$.\footnote{The original form of Eq. \ref{su2^N}
which appeared in the first version of this paper was not general enough
to correctly reproduce the curves for $N\geq 4$. This was noted by
G.~Hailu in~\cite{Hailu}.}

%%%%%%%%%%%%%%%%%%%%%%%%%%%%%%%%%
\section{$SU(N)\times SU(N)$\label{sec:SUN2}}
%%%%%%%%%%%%%%%%%%%%%%%%%%%%%%%%%

Next we generalize 
the $SU(2)\times SU(2)$ theory  presented in Section~\ref{sec:SU2N}
to $SU(N)\times SU(N)$ with
fields $Q_1$ and $Q_2$ transforming as $(\Yfund,\overline{\Yfund})$ and
$(\overline{\Yfund},\Yfund)$.
Along generic flat directions $SU(N)\times SU(N)$ is broken to $U(1)^{N-1}$,
as shown in the appendix.
Therefore this theory is in the Coulomb phase.
Since there is a non-anomalous $U(1)_R$ symmetry under which the
fields $Q_1$ and $Q_2$ have R-charge zero, there can be
no dynamical superpotential generated; thus, the Coulomb
phase is not lifted. 

The independent gauge invariant operators are $B_1= {\rm det}\,Q_1$, 
 $B_2={\rm det}\,Q_2,$ and 
$T_n={\rm Tr} (Q_1Q_2)^n$, $n=1,
\cdots, N-1$.
  This agrees
with the counting of degrees of freedom:  the fields
$Q_1$ and $Q_2$ contain
$2N^2$ complex degrees of freedom and there are $2(N^2-1)$ 
$D$-flat conditions.  Since there is an unbroken 
$U(1)^{N-1}$ gauge symmetry only $2(N^2-1)-(N-1)$ of these
conditions are independent and thus one expects to find
$2N^2-[2(N^2-1)-(N-1)]=(N+1)$ independent gauge
invariant objects, which exactly matches the number
of operators listed above.  
We again assume that 
there is a hyperelliptic curve describing this
theory involving these degrees of freedom and the scales
$\Lambda_1$ and $\Lambda_2$.

We will present the $SU(3)\times SU(3)$ case in detail 
and then generalize to $SU(N)\times SU(N)$. The matter field content of the
$SU(3)\times SU(3)$ theory is
\begin{displaymath}
\begin{array}{c|cc}
& SU(3) & SU(3) \\ \hline
Q_1 & \Yfund & \overline{\Yfund} \\
Q_2 & \overline{\Yfund} & \Yfund \end{array}
\end{displaymath}
The independent gauge invariants are
\begin{eqnarray}
&B_1= {\rm det}\, Q_1,\; \; &B_2= {\rm det}\, Q_2, \nonumber \\
&T_1= {\rm Tr}\, Q_1Q_2,\; \; & T_2= {\rm Tr}\, (Q_1Q_2)^2.\nonumber
\end{eqnarray}
All other gauge invariants can be expressed in terms of these four.  For
example, the operator $T_3={\rm Tr}\, (Q_1Q_2)^3$
is constrained classically via the identity \begin{equation}
{\rm det}\,M=B_1\,B_2,\label{eq:cl-constraint} \nonumber \end{equation}
where $M_\alpha^\beta=Q_{1\;A}^{\,\alpha}\,Q_{2\;\beta}^{\,A}$.
To see this we first express ${\rm det}\, M$ in terms of the invariants
$T_i$, $i=1,2,3$ as
\begin{eqnarray}
{\rm det}\,M=\frac{1}{6}\left(T_1^3-3T_1T_2+2T_3\right).
\end{eqnarray}
The classical constraint of Eq.~\ref{eq:cl-constraint} then 
yields\begin{equation}
T_3^{\rm cl}=\frac{1}{2}\left(6B_1B_2+3T_2T_1-T_1^3\right).
\label{eq:T3cl}\end{equation}
It is natural to consider the composite field $\Phi^\alpha_\beta=
Q_{1\,A}^\alpha Q_{2\,\beta}^A-\frac{1}{3}{\rm Tr}\,Q_1Q_2\delta^\alpha_\beta$,
which is a singlet
under one of the $SU(3)$'s and an adjoint under the other.  We define 
\begin{eqnarray}
u&=&\frac{1}{2}{\rm Tr}\,\Phi^2 = \frac{1}{2}\left(T_2-\frac{1}{3}T_1^2\right),
 \nonumber \\
v&=&\frac{1}{3}{\rm Tr}\,\Phi^3 =\frac{1}{3}\left(T_3^{cl}-T_2T_1+
\frac{2}{9}T_1^3\right) \nonumber \\
 &=&\frac{1}{3}\left(3B_1B_2+\frac{1}{2}T_2T_1-\frac{5}{18}T_1^3\right)\, ,
\label{eq:uv} \end{eqnarray}
which correspond to the moduli of an $SU(3)$ theory with adjoint field $\Phi$.
It turns out that the $SU(3)\times SU(3)$ curve depends only on these
combinations of $T_i$ and $B_i$.

As there are generically two $U(1)$'s unbroken, 
we expect there to be a
genus two hyperelliptic curve describing the theory, given by a sixth order
polynomial in $x$.  Having identified the moduli space 
we consider various limits to 
determine the coefficients of this hyperelliptic curve.
Consider the limit where $Q_1$ gets a large diagonal VEV, $w$, 
$w\gg\Lambda_1,\Lambda_2$. 
Then $SU(3)\times SU(3)$ is broken to the diagonal
$SU(3)_D$.  Under $SU(3)_D$, $Q_1$ and $Q_2$ decompose into two singlets and
two adjoints.  The adjoint from $Q_1$ is eaten, leaving two
singlets, which are assumed not to
enter the gauge dynamics, and an adjoint, $\Phi_D=Q_2-\frac{1}{3}{\rm Tr}
\,Q_2$.  The scale of the resulting $SU(3)_D$ theory is determined by matching
at the scale $w$ which gives $\Lambda_D^6=\Lambda_1^6\Lambda_2^6/w^6=
\Lambda_1^6\Lambda_2^6/B_1^2$.  The dynamics of this effective $N=2,\,SU(3)$
gauge theory is described by the curve
\begin{equation} 
y^2=(x^3-u_Dx-v_D)^2-4\Lambda_D^6, 
\label{eq:3.2}
\end{equation}
and the invariant traces $u_D,\,v_D$ can easily be expressed in terms of $u,
\,v$ of Eq.~\ref{eq:uv},
\begin{eqnarray}
u_D &=&\frac{1}{2}{\rm Tr}\,\Phi_D^2
%=\frac{1}{2}\left({\rm Tr}\,Q_2^2-
%\frac{1}{3}({\rm Tr}\,Q_2)^2\right) \nonumber \\
  =\frac{1}{2B_1^{2/3}}\left(T_2-\frac{1}{3}T_1^2\right)=
\frac{u}{B_1^{2/3}} \nonumber \\
v_D&=&\frac{1}{3}{\rm Tr}\,\Phi_D^3=\frac{1}{3B_1}\left(T_3^{\rm cl}
-T_2T_1+\frac{2}{9}T_1^3\right)= \frac{v}{B_1}
\end{eqnarray}

The curve in Eq.~\ref{eq:3.2} can then be written in terms of the original 
$SU(3)\times SU(3)$ gauge invariants and the original 
scales: 
\begin{equation}
y^2=\left(x^3-\frac{u}{B_1^{2/3}}\,x-\frac
{v}{B_1}\right)^2-
4\frac{\Lambda_1^6\Lambda_2^6}{B_1^2}. 
\end{equation}
Rescaling $x\rightarrow x/B_1^{1/3}, y\rightarrow y/B_1$, 
the curve takes the form
\begin{equation} 
y^2=\left(x^3-ux-v\right)^2-4\Lambda_1^6\Lambda_2^6. 
\label{eq:3.6}
\end{equation}

%%%%%%%%%%%%%%%%
The hyperelliptic curve of the $SU(3)\times SU(3)$ theory must reproduce 
Eq.~\ref{eq:3.6} in the limit of large diagonal VEV $w$ for $Q_1$, but the
sixth order polynomial which describes it may well contain new terms which
are not yet fixed because they are subdominant in this limit.  We now write
a more general polynomial, containing all terms consistent with the 
$R$-symmetry of the theory, and the assumption that the scales appear only as
integer powers of $\Lambda_1^6$ and $\Lambda_2^6$, corresponding to instanton
effects.  The theory has an anomalous $U(1)_R$ symmetry in which $Q_1,\,Q_2,\,
\Lambda_1$, and $\Lambda_2$ have $R$-charge one.  Covariance of the curve in
the large VEV limit (Eq.~\ref{eq:3.6}) requires that $x$ and $y$ be
assigned $R$-charges two and six, respectively.  These $R$-charge assignments
are summarized in the table below.

\begin{center}\begin{tabular}{l|c|c|c|c|c|c}
& $y$ & $x$ & $T_1$ & $T_2$ & $B_1, B_2$ & $\Lambda_1^6,\Lambda_2^6$ \\ \hline
$U(1)_R$& 6 & 2 & 2 & 4 & 3 & 6  \end{tabular}\end{center}

The most general sixth order polynomial including all terms consistent
with these requirements and the discrete $\Lambda_1\leftrightarrow\Lambda_2$
symmetry is
\begin{eqnarray}
&y^2=x^6-2ux^4-\left(2v+\alpha\left(\Lambda_1^6+\Lambda_2^6\right)\right)
x^3+u^2x^2+\left(2uv+\beta\left(\Lambda_1^6+\Lambda_2^6\right)u\right)x
&\nonumber \\ &
+v^2-\gamma\Lambda_1^6\Lambda_2^6+\delta\left(\Lambda_1^6+\Lambda_2^6
\right)^2 +\epsilon\left(\Lambda_1^6+\Lambda_2^6\right)v\; ,&
\label{eq:SU3D}\end{eqnarray}
with as yet undetermined coefficients $\alpha,\beta,\gamma,\delta,\epsilon.$
Additional terms involving other combinations of products 
of the fields and the scales are not consistent with the
large VEV limit.  Other combinations of gauge invariants and scales are
excluded by the strong coupling limit, which we now describe.
        
Next we consider the limit where $SU(3)_2$ is strong, $\Lambda_2\gg\Lambda_1$.
$Q_1$ and $Q_2$ confine to form three singlets under the remaining $SU(3)_1$, 
$Q_1^3,
Q_2^3$, ${\rm Tr}\,Q_1Q_2$, and an adjoint
$\Phi_1=\frac{1}{\mu}(Q_1Q_2-{\rm Tr} \, Q_1Q_2)$, where the scale $\mu$
is introduced to give the adjoint canonical dimension one.  Below the scale
$\Lambda_2$ we have an
$SU(3)$ theory with an adjoint and scale $\Lambda_1$.

The confining $SU(3)_2$ theory has a quantum modified 
constraint~\cite{Seiberg}
\begin{equation}
{\rm det}\,M-B_1B_2=\Lambda_2^6.\nonumber  
\end{equation}
This quantum modified constraint will result in the 
expression (\ref{eq:T3cl}) for $T_3$ being modified by
the addition of $3\Lambda_2^6$.

We identify the moduli in this limit, 
\begin{eqnarray}
u_1&=&\frac{1}{2\mu^2}\left(T_2-\frac{1}{3}T_1^2\right) =\frac{u}{\mu^2} 
\nonumber \\
v_1&=&\frac{1}{3\mu^3}\left(T_3-T_2T_1+\frac{2}{9}T_1^3\right) 
=\frac{1}{3\mu^3}\left(3B_1B_2+\frac{1}{2}T_2T_1-\frac{5}{18}T_1^3+
3\Lambda_2^6\right) \nonumber \\
 &=&\frac{1}{\mu^3}\left(v+\Lambda_2^6\right)\, .
\end{eqnarray}
The curve in this limit is then, after rescaling $x\rightarrow x/\mu, y
\rightarrow y/\mu^3$,
\begin{equation}
y^2=\left[x^3-ux-v-\Lambda_2^6\right]^2-4\mu^6\Lambda_1^6. 
\end{equation}

This fixes the previously undetermined parameters in Eq.~\ref{eq:SU3D} except
for $\gamma$.
At this stage, using the $\Lambda_1\leftrightarrow \Lambda_2$ flavor
symmetry and the above limits, the $SU(3)\times SU(3)$ curve takes the form
\begin{equation}
y^2=\Big{[}x^3-ux-v-\Lambda_1^6-\Lambda_2^6\Big{]}^2-
\gamma\Lambda_1^6\Lambda_2^6.
\label{eq:t+g}
\end{equation}

In order to determine the coefficient $\gamma$ we higgs the 
theory to
$SU(2)\times SU(2)$.  Consider the limit where $Q_1$ and $Q_2$ each get large 
VEVs of the form
\begin{equation}
Q_1=Q_2=\left(\begin{array}{ccc}
w & & \\ 
 & 0 & \\
 & & 0
\end{array} \right).
\label{eq:w}\end{equation}
Then $SU(3)\times SU(3)$ is broken to $SU(2)\times SU(2)\times U(1)$ with
the uneaten degrees of freedom lying in the
two by two lower right block  of the fields
$Q_1$ and $Q_2$, which we denote by $q_{\,\alpha}^{\,A}$ and 
$\tilde{q}_{\,A}^{\,\alpha}$.  
These remaining degrees of freedom are neutral under the $U(1)$.

The non-perturbative description of this higgs limit of the curve 
Eq.~\ref{eq:t+g} is the following:  Given values of the invariants $u$ and
$v$, Newton's formula of Eq.~\ref{eq:newton} can be used to find $s_2$ and
$s_3$ and then a set of values for $a_1,\,a_2,\,a_3$ with $\sum a_i=0$ via
Eq.~\ref{eq:s=a}.  Note that in the strong coupling regime the $a_i$'s are
not the VEVs of any fundamental field, although classically they are the
diagonal VEVs of the composite adjoint field $\Phi=Q_2Q_3-\frac{1}{3}{\rm Tr}
\,Q_2Q_3$.  The curve in Eq.~\ref{eq:t+g} can be rewritten as
\begin{equation}
y^2=\left(\prod_{i=1}^3 (x-a_i)-\Lambda_1^6-\Lambda_2^6\right)^2
-\gamma\Lambda_1^6\Lambda_2^6\, .
\label{eq:t+g2}\end{equation}
Without loss of generality we can take $a_3=\frac{2}{3}\rho,\,a_1=-\frac{1}{3}
\rho+a,\,a_2=-\frac{1}{3}\rho-a$.  We shift $x\rightarrow x-\frac{1}{2}\rho$,
and reexpress the curve as
\begin{equation}
y^2=\left((x-\rho)(x-a)(x+a)-\Lambda_1^6-\Lambda_2^6\right)^2-\gamma
\Lambda_1^6\Lambda_2^6.
\label{eq:rho}\end{equation}
The higgs limit is  
$\rho\gg a$ in which one pair of branch points recedes
toward infinity, as in \cite{Argyres}, and monodromies and periods calculated 
in the finite
region are those of the $SU(2)\times SU(2)$ theory.  To see this concretely
we rescale $y\rightarrow y(x-\rho)$, and assume $x\ll \rho$.  In this region
the curve Eq.~\ref{eq:rho} may be expressed as the approximate genus one
curve \begin{equation}
y^2=\left(x^2-a^2+\frac{\Lambda_1^6}{\rho}+\frac{\Lambda_2^6}{\rho}\right)^2-
\gamma\,\frac{\Lambda_1^6\Lambda_2^6}{\rho^2}. \end{equation}
This agrees with Eq.~\ref{su2su2} if we identify $a^2=U,\,\gamma=4$ and
scales $\rho\tilde{\Lambda_i^4}=\Lambda_i^6$.  In the classical region, it
is certainly the case that $a^2=U$ and $\rho=w^2$ of Eq.~\ref{eq:w}.  The
scales are then related by the standard matching condition.  We have now
completely determined the $SU(3)\times SU(3)$ curve,\begin{equation}
\label{su3su3curve}
y^2=\left(x^3-ux-v-\Lambda_1^6-\Lambda_2^6\right)^2
-4\Lambda_1^6\Lambda_2^6\,.
\end{equation}

The generalization of this analysis to $SU(N)\times SU(N)$ is 
straightforward.
The effect of the quantum modified constraint is to shift the
classical expression for $s_N$ by $(-1)^N(\Lambda_1^{2N}+
\Lambda_2^{2N})$.  Recall
the curve of Eq.~\ref{suncurve} for the $SU(N)$ theory with an adjoint,
\begin{equation}
  y^2=\left(\sum_{i=0}^Ns_ix^{N-i}\right)^2-4\Lambda^{2N}. \nonumber 
\end{equation}
The previous arguments carry
through in direct analogy, resulting in the curve for the $SU(N)\times SU(N)$ 
theory:
\begin{equation}
 y^2=\left(\sum_{i=0}^Ns_ix^{N-i}+(-1)^N\left(\Lambda_1^{2N}+\Lambda_2^{2N}
 \right)\right)^2-4\Lambda_1^{2N}\Lambda_2^{2N},
\end{equation}
where the $s_i$'s are the symmetric invariants of the
composite adjoint $\Phi=Q_1Q_2-\frac{1}{N}{\rm Tr}\,Q_1Q_2$
and are to be expressed in terms of the gauge invariants $T_i$ 
and $B_i$ via classical expressions.

As a consistency check on the $SU(3)\times SU(3)$ curves
we consider integrating out the fields $Q_1$ and $Q_2$ by adding
a mass term 
\[ W_{tree}=mT_1 \]
to the superpotential. Then the low-energy theory will be a pure 
$SU(3)\times SU(3)$ Yang-Mills theory and we expect to find
nine vacuum states.

The effective low-energy superpotential has to account
for the monopoles and dyons which become massless along
the singular surfaces of the hyperelliptic curve of 
Eq.~\ref{su3su3curve}. 
These singular surfaces can be determined 
by finding the zeros of the discriminant $\Delta$ of the 
curve. For the $N=2$ $SU(3)$ curve described by
$y^2=(x^3-ux-v)^2-4\Lambda^6$,
the discriminant factorizes, $\Delta \propto \Delta_{+}
\Delta_{-}$, where 
$\Delta_{\pm}=4u^3-27(v\pm 2\Lambda^3)^2$~\cite{ArgyresKlemm}. 
In our
case $u$ and $v$ are expressed in terms of $T_1$, $T_2$, $B_1$ and
$B_2$ by Eq.~\ref{eq:uv}.
Thus the effective superpotential can be written as
\[ W= \Delta_{+} \tilde{E}_{+}E_{+} + \Delta_{-}\tilde{E}_{-}
E_{-}+m T_1,\]
where the $E_{+},\tilde{E}_{+}$ fields correspond to the
monopoles which become massless at $\Delta_{+}=0$ and the
 $E_{-},\tilde{E}_{-}$ fields correspond to the
dyons which become massless at $\Delta_{-}=0$.
The $T_1$ equation of motion will force at least
one of the monopole condensates to be non-vanishing. But
then the $T_2$ equation will force both $\tilde{E}_{+}E_{+}$
and $\tilde{E}_{-}E_{-}$ to be non-zero, which by the 
$\tilde{E}_{+},\tilde{E}_{-}$ equations lock the fields
to one of the $Z_3$ symmetric singularities $\Delta_{+}=
\Delta_{-}=0$. The $B_1$ and $B_2$ equations just set 
$B_1$ and $B_2$ to zero, while  $\tilde{E}_{+}E_{+}$
and  $\tilde{E}_{-}E_{-}$ can be uniquely determined once 
$T_1$ and $T_2$ are fixed. Thus in order to count the
number of vacua one needs to solve the equations 
$\Delta_{+}=\Delta_{-}=B_1=B_2=0$ for the variables $T_1$ and
$T_2$. Using Eq.~\ref{eq:uv} with $B_1=B_2=0$ these can be written as
\begin{eqnarray}
&& \frac{1}{2}T_2T_1- \frac{5}{18}T_1^3-3\Lambda_1^6-
3\Lambda_2^6=0 \nonumber \\
&& T_2-\frac{1}{3}T_1^2= 3\omega \Lambda_1^2\Lambda_2^2, \nonumber
\end{eqnarray}
where $\omega$ is a third root of unity.
One can see that for each value of $\omega$ we get a cubic
equation for $T_1$, therefore we conclude that there are 
nine distinct vacua in agreement with the Witten index.
We do not find detailed agreement with the vacua determined by the original
integrating in procedure of \cite{intin}, which we would expect to be at 
$T_2=B_i=0,\,T_1=\omega_1
\Lambda_1^2+\omega\Lambda_2^2$.  However, there are other examples, such as
the $N=2\,,SU(N_c)$ gauge theory with $N_c>4$, where the operator $u_{2k}$ 
can mix with $u_k^{\,2}$, for example, in which the naive integrating 
in procedure does not reproduce the VEV's of the gauge invariant operators
found from the effective superpotential including massless monopoles and
dyons.  Similarly, in the $SU(3)\times SU(3)$ theory, $T_2$ and $T_1^{\,2}$
can mix, and we should not expect the naive integrating in procedure to
work.\footnote{We thank Ken Intriligator for pointing this out to us.
See also Ref.~\cite{Israel}.}

%%%%%%%%%%%%%%%%%%%%%%%%%%%
\section{$SU(N)^M$}
\label{sec:SUNM}
%%%%%%%%%%%%%%%%%%%%%%%%%%%
Finally, we generalize the previous analysis to 
$SU(N)^M$ theories with matter content given below:
\begin{equation}
 \begin{array}{c|ccccc}
      & SU(N)_1 & SU(N)_2 & SU(N)_3 & \cdots & SU(N)_M \\ \hline
  Q_1 & \Yfund  & \overline{\Yfund}  & 1       & \cdots & 1 \\
  Q_2 & 1       & \Yfund  & \overline{\Yfund}  & \cdots & 1 \\
  \vdots & \vdots & \vdots & \vdots & \vdots & \vdots \\
  Q_M & \overline{\Yfund} & 1 & 1 & \cdots & \Yfund 
 \end{array} \nonumber
\end{equation}
The independent gauge invariants are $B_i={\rm det}\,Q_i,\, i=1,\dots,M$; 
and $T_i={\rm Tr}\,(Q_1\cdots Q_M)^i,\,i=1,\dots,N-1$.
We define the composite field 
\begin{displaymath}
  \Phi=Q_1 Q_2\cdots Q_M - \frac{1}{N} {\rm Tr}\, Q_1 Q_2\cdots Q_M,
\end{displaymath}
which is an adjoint under one of the 
$SU(N)$'s and invariant
under the others. From $\Phi$ we form the invariants 
$s_i$ as in Section~\ref{sec:SU2N}. These invariants $s_i$,
when expressed in terms of the invariants $T_i$ and $B_i$, classically have
the same functional 
form as in the $SU(N)\times SU(N)$ case, except that $B_1B_2$ has to be
replaced by 
the product
over all the $B_i$'s.
 
In terms of these variables, the $SU(N)^M$ curve is given by
\begin{equation}
y^2=\left[\sum_{i=1}^N s_i(T_j,B_j)\,x^{N-i}+ (B_iB_{i+1}\rightarrow \Lambda_{i+1}^{2N})\right]^2-
4\prod_{j=1}^M \Lambda_i^{2N}. \label{eq:SU(N)^M} 
\end{equation}
The modulus $s_N$ is written in terms of $T_i$ and $B_i$ via classical relations, and
contains the term $(-1)^N\prod_i B_i$, for example.  
In Eq.~\ref{eq:SU(N)^M} the second term in brackets is shorthand for replacing sets of
nearest neighbor bilinears $B_iB_{i+1}$ by $\Lambda_{i+1}^{2N}$.\footnote{As for the 
$SU(2)^N$ theory, some terms were 
missing from Eq. \ref{eq:SU(N)^M} in a previous version.}  For example, we would
substitute $B_1B_2B_3B_4$ by
$B_1B_2B_3B_4+\Lambda_1^{2N} B_2B_3
+\Lambda_2^{2N} B_3B_4+\Lambda_3^{2N} B_4B_1+\Lambda_4^{2N} B_1B_2 
+\Lambda_1^{2N}\Lambda_3^{2N}
+\Lambda_2^{2N}\Lambda_3^{2N}$.
We can check that this curve produces the correct $SU(N)^{M-1}$ curve 
upon higgsing
the theory and in the strong $SU(N)_M$ limit.
First consider breaking to $SU(N)^{M-1}$ by giving the
field $Q_M$ a large diagonal VEV $w$.  In this limit the degrees of freedom
are $B_i/w,\,i=1,\dots,M-1$; and $T_j/w^j,\,j=1,\dots,N-1$.  
Upon rescaling the gauge invariants by appropriate powers of the VEV
$w$ and using the matching relation $w^{2N}\Lambda_D^{2N}=B_M\Lambda_D^{2N}=
\Lambda_{M-1}^{2N}\Lambda_M^{2N}$ the curve reproduces the correct
$SU(N)^{M-1}$ limit. 

We can also check the curve in the strong $SU(N)_M$ limit.  In this limit
$SU(N)_M$ confines with a quantum modified constraint and we obtain
another $SU(N)^{M-1}$ theory in this limit. 
The degrees of freedom are 
\begin{eqnarray}
& & B_i, \; i=1,\dots,M-2; \; \; T_n, n=1,\dots,N-1; \nonumber \\
&& {\rm and}  \; \; 
\tilde{B}_{M-1}
=(\frac{1}{\mu}Q_{M-1}Q_M)^N=\frac{1}{\mu^N}(B_{M-1}B_M+
\Lambda_M^{2N}). \nonumber
\end{eqnarray}
The scale $\mu$ is introduced as usual to give the field $Q_{M-1}Q_M$
canonical dimension one.  Comparing the curves for $SU(N)^{M}$
and $SU(N)^{M-1}$ with
these degrees of freedom fixes the scale $\mu=\Lambda_M$, and then 
Eq.~\ref{eq:SU(N)^M} agrees with the curve for the $SU(N)^M$ theory.

Similarly, we can consider higgsing the theory to $SU(N-1)^M$ by giving all
the $Q_i$'s a VEV in one component, in which case we again find agreement
amongst the curves.  

%%%%%%%%%%%%%%%%%%%%%%%%%%%%%%%%%%%%%%%%%%%%%
\section{Conclusions\label{sec:conclusions}}  
%%%%%%%%%%%%%%%%%%%%%%%%%%%%%%%%%%%%%%%%%%%%%

    We have extended the results of Ref.~\cite{phases} 
to $N=1$ supersymmetric  gauge theories with
product gauge groups $SU(N)^M$ and $M$ chiral superfields 
in the fundamental representation of exactly two of the
$SU(N)$ factors. These theories have an unbroken $U(1)^{N-1}$
gauge group along generic flat directions and are therefore
in the  Coulomb phase. For
$M=2$ there are two limits in which the low-energy degrees of 
freedom are
those of an effective $N$=2 $SU(N)$ 
gauge theory, so it is natural to assume that
the gauge kinetic functions are given in general by the period matrix of genus 
$N-1$ hyperelliptic curves.  We then derive those curves by 
studying these two limits.
For $M>2$ the
$SU(N)^{M-1}$ theory can be obtained by higgsing the
$SU(N)^M$ theory, so we assume again that 
hyperelliptic curves determine the Coulomb phase dynamics and 
then find the curves by studying limits.

    There is a systematic pattern of curves for the $SU(N)\times 
SU(N)$ models. When 
written in terms of trace of powers of the composite 
adjoint field $Q_1 Q_2$, the new curves are related
to the known~\cite{ArgyresKlemm} curves for $N=2$ $SU(N)$
 theories by a simple shift due to the
quantum modified constraint in the product group models. 
For $M>2$ the
curves are entirely new, and they depend on the $N+M-1$ invariants in a complex
but systematic way.

    One of the most striking aspects of the work 
of~\cite{SW,phases} is that one can
add a mass term to the original theory and demonstrate that 
magnetic
confinement occurs. We have studied this mechanism in our 
$SU(2)^3$ and $SU(3)^2$
models. In the first case we find 8 confining vacua with detailed agreement 
between the two approaches based on the low-energy superpotential with
monopole fields and the dynamical superpotential describing gaugino 
condensation after integrating out massive chiral fields. In the second case
we find, as expected, 9 vacua from the low-energy monopole superpotential as 
well.  For $N>2$, $M>2$ our models are chiral, and they
are the first examples of chiral theories in the Coulomb phase. 

\section*{Acknowledgments}
 We are extremely grateful to Ken Intriligator and Nathan 
Seiberg for several useful discussions.
D. F. is supported in part by the NSF grant PHY-92-06867.  
C. C., J. E. and W. S. are supported in part by the DOE under 
cooperative agreement DE-FC02-94ER40818.

%%%%%%%%%%%%%%%%%%%%%%%%%%%%%%%%%%%%%%%%%%%%%%%%%%%%%%%%%%%%%%%%%%%%%%%%%%%%
\appendix
\section*{Appendix: D-flat Conditions in $SU(N)^{M}$ theories}
%%%%%%%%%%%%%%%%%%%%%%%%%%%%%%%%%%%%%%%%%%%%%%%%%%%%%%%%%%%%%%%%%%%%%%%%%%%%
In the early approach to the dynamics of SUSY gauge theories~\cite{ADS}
the moduli space of SUSY vacua was
obtained by finding the most general field configuration, up to
symmetries, which satisfies the D-flat conditions. In the more
modern approach~\cite{Seiberg}, which emphasizes holomorphy, the moduli
space is parameterized by a set of algebraically independent
gauge-invariant polynomial functions of the chiral superfields~\cite{Markus}.
Such a set of $N+1$ polynomial invariants were specified
in Section~\ref{sec:SUN2} for the $SU(N)^2$ models, and a set of $N + M - 1$
invariants for the $SU(N)^{M}$ models was given in Section~\ref{sec:SUNM}.
It is a simple but useful consistency check to study the D-flat
conditions. This will confirm the dimension of the moduli space
and make manifest the residual $U(1)^{N-1}$ symmetry of the
Coulomb phase.

We begin with the $SU(N) \times SU(N)$ case. As a simple variant
of a standard matrix representation we know that $Q_1$ and $Q_2$
can be expressed in the form
\begin{equation}
  \label{eq:A-1}
  Q_1 = U_1\, A\, V^{-1}_{2} \, e^{i\alpha},\;\;
  Q_2 = U_2\, B\, V^{-1}_{1} \, e^{i \beta},
\end{equation}
where $U_i$ and $V_i$ are $SU(N)$ matrices and $A$ and $B$ are
real diagonal matrices whose diagonal elements $a_i$ and $b_i$
are ordered, i.e. $a_1 > a_2 > \,\ldots\, > a_N > 0$.
We assume a generic configuration in which none of the $a_i$
coincide, and the same for the $b_i$. The $U_i$ and $V_i$ are
still not uniquely determined; there remains the freedom $U_1 \to
U_1C$, $V_2 \to V_2C$ and $U_2 \to U_2C^{\prime}$, $V_1 \to
V_1C^{\prime}$ where $C$ and $C^{\prime}$ are diagonal $SU(N)$
matrices, but this redundancy plays little role in the subsequent
analysis.

The original form of the D-flat conditions is
\begin{eqnarray}
  \label{eq:A-2}
  D^a_1 &=& {\rm Tr}\,(Q^{\dagger}_1\, T^a \, Q_1 \, - \, Q_2\, T^a\, 
Q_2^{\dagger}) = 0, \\
  D^a_2 &=& {\rm Tr}\,(Q^{\dagger}_2\, T^a\, Q_2 \, - \, Q_1\, T^a \, 
Q^{\dagger}_1) = 0.
  \nonumber
\end{eqnarray}
Since the generators $T^a$ of the fundamental representation of
$SU(N)$ are a complete set of $N \times N$ Hermitian matrices,
this means that $Q_1 Q^{\dagger}_1 \, -\, Q^{\dagger}_2 Q_2$ and
$Q_2 Q^{\dagger}_2 \, - \, Q^{\dagger}_1 Q_1$ must be multiples
of the identity. Using Eq.~\ref{eq:A-1}, these conditions read
\begin{eqnarray}
  \label{eq:A-3}
  U_1\, A^2\, U_1^{-1} \, -\, V_1\, B^2\, V^{-1}_1 & = & c_1
  \bf{1}, \nonumber \\
  V_2\, A^2\, V^{-1}_2 \, - \,  U_2 \, B^2 \, U^{-1}_2 & = & c_2
  \bf{1}, 
\end{eqnarray}
where $c_1$ and $c_2$ are real constants. Taking traces gives $c_1
= c_2 = c$. Suppose one brings the $B^2$ term to the right side of
the first equation. One can then see that the characteristic
equations for the matrices $U_1\, A^2\, U^{-1}_1$ and $V_1 (B^2
\, - \, c_1 {\bf 1})\, V^{-1}_1$ must have the same
roots. This means that $a^2_i = b^2_i + c$ for each diagonal
element of $A$ and $B$.

Conjugate the first equation in \ref{eq:A-3} by $U^{-1}_1 \,
\ldots\, U_1$ and the second by $V_2^{-1} \, \ldots\,
V_2$. Rearrange the resulting expressions to read 
\begin{equation}
  \label{eq:A-4}
  \left[ U^{-1}_1 V_1\,, \, B^2 \right] = 0 \; \; {\rm and}\; \;
   \left[V^{-1}_2 U_2\,, \, B^2 \right] = 0\,.
\end{equation}
These imply that $U^{-1}_1 V_1 = C_1$ and $V_2^{-1} U_2 = C_2$
where $C_1$ and $C_2$ are diagonal $SU(N)$ matrices. We can now
return to (\ref{eq:A-1}) and rewrite it as
\begin{equation}
  \label{eq:A-5}
  Q_1 = U_1\, A\, C_2\, U^{-1}_2 e^{i \alpha} \; \; {\rm and}\; \;
  Q_2 = U_2\, B\,C^{-1}_1 U^{-1}_{1} e^{i \beta}\,.\nonumber
\end{equation}

Now make an $SU(N)_1$ gauge transformation by $U^{-1}_1$, and an
$SU(N)_2$ transformation by $C_2U^{-1}_2$ to obtain the canonical
diagonal representation
\begin{equation}
  \label{eq:A-6}
  Q_1 = 
  \left(
    \begin{array}{llll}
a_1\\
& a_2\\
&&\ddots\\
&&&a_N
    \end{array}
      \right)\, e^{i \alpha}, \quad
%%%%%%%%%%%%%
Q_2 =
\left(
  \begin{array}{llll}
b_1\\
& b_2\\
&&\ddots\\
&&&b_N
  \end{array}
\right)\, C\ e^{i\beta}.
\end{equation}
with $C= C_2\, C^{-1}_1$ and $a_i = (b^2_i + c)^{1/2}$. There are
$N$ real variables and $N$ phases in $Q_2$, and the additional
real $c$ and phase $\alpha$. This is the correct count of
independent variables of the Coulomb phase. Further one sees that
the gauge is not completely fixed because the canonical
representation is invariant under $Q_1 \to C^{\prime}\, Q_1\,
C^{\prime - 1}$ and $Q_2 \to C^{\prime}\, Q_2\, C^{\prime -1}$ where
$C^{\prime}$ is a diagonal matrix of the diagonal $SU(N)$
subgroup. This is just the expected residual $U(1)^{N-1}$ gauge
invariance of an $N-1$ dimensional Coulomb branch.

This approach can be extended to the $SU(N)^{m}$ case with chiral
superfields in the $(\Yfund ,\overline{\Yfund})$
 representation of $SU(N)_{i} \times
SU(N)_{ i+1}$, and otherwise inert, denoted by $Q_{ij}$. (We
always take $j = i + 1$ for $ i = 1 \, \ldots\, m-1$, and $j = 1$
for $i= m$.) We start with representations
\begin{equation}
  \label{eq:A-7}
  Q_{ij} = U_{i} A_{ij} V_{j}^{-1} e^{i \alpha_{ij}},
\end{equation}
where $U_i$ and $V_j \in SU(N)$ and $A_{ij}$ is a real diagonal
matrix.  $\*$As before there is a non-uniqueness $U_i \to U_i\,
C_{ij}$, $V_j \to V_j\, C_{ij}$ with $C_{ij}$ diagonal and 
${\rm det}\,C_{ij} = 1$.

There are $m$ independent D-flat conditions, and one learns, as in
(\ref{eq:A-3}), that the $SU(N)_j$ condition implies
\begin{equation}
  \label{eq:A-8}
  U_j\, A^2_{j k}\, U^{-1}_j \, - \, V_j\, A^2_{ij}\, V^{-1}_j =
  c_j \bf{1}.
\end{equation}
(Again we take $j = i + 1$, and $k = j + 1$ with wraparound where
required.) The sum of traces of these $j$ equations just gives
$\sum^m_{j= 1}c_j = 0$, and one finds from considering
characteristic equations that diagonal elements $a_{ij,\rho}$,
satisfy $a_{jk,\rho}^2 = a_{ij, \rho}^2 + c_j$, for $\rho = 1, 2
\, \ldots\, N$.

By appropriate conjugation and rearrangement, one again finds
that $U_i^{-1} \, V_i = C^{-1}_i$, so that Eq.~\ref{eq:A-7} becomes
\begin{equation}
  \label{eq:A-9}
  Q_{ij} = U_i A_{ij} C_j U^{-1}_j e^{i \alpha_{ij}}.
\end{equation}
We now make a gauge transformation by $U_i$ in each $SU(N)_i$
factor group which brings us to the diagonal representation
\begin{equation}
  \label{eq:A-10}
  Q_{ij} = A_{ij} \, C_j\, e^{i \alpha_{ij}}.
\end{equation}
The final $SU(N)_1 \times SU(N)_2 \times SU(N)_m$ gauge transform
by $(1, C_2, C_2C_3,\, \ldots$, $C_2 C_3 \, \ldots\,
C_m)$ then gives the canonical representation
\begin{eqnarray}
  \label{eq:A-11}
  Q_{ij} &=& A_{ij} e^{i \alpha_{ij}}, \qquad i = 1 \, \ldots \,
  m-1, \\
  Q_{m1} &=& A_{m1} C\, e^{i\alpha_{m1}}, \nonumber
\end{eqnarray}
with $C = C_1C_2 \, \ldots\, C_m$. This contains $N + m -1$
independent real variables and $m + N -1$ phases in agreement
with the number of independent holomorphic polynomials found in
Section~\ref{sec:SUNM}. There is again a residual unfixed $U(1)^{N-1}$
Coulomb phase gauge symmetry.

\end{document}